\documentclass[12pt]{article}
\usepackage{epsf}
\usepackage{epsfig}

\usepackage{amssymb}
\usepackage{geometry}
\usepackage{amsmath}
\usepackage{amsbsy}
\usepackage{amstext}
\usepackage{amsfonts}
\usepackage{tabularx}
\usepackage{hhline}

\newcommand{\projectorl}{\text{P}_\text{L}}
\newcommand{\projectorr}{\text{P}_\text{R}}

\begin{document}

\begin{titlepage}
\begin{flushright}
10 March 2005
\end{flushright}
\vskip1.5cm

\begin{center}
\boldmath
{\Large\bf Spontaneous $CP$ Symmetry Breaking\\[0.3cm]
 at the Electroweak Scale}
\unboldmath
\vskip 2cm Cristi\'{a}n Valenzuela
\vskip .5cm

\vspace{0.9cm}
{\it Institut f\"ur Theoretische Physik E, RWTH Aachen\\
D--52056 Aachen, Germany\\}
\vspace{0.3cm}
{\it Departmento de F\'{\i}sica,
Universidad T\'{e}cnica Federico Santa Mar\'{\i}a\\
Valpara\'{\i}so, Chile\\}
\vspace{3\baselineskip}

\vspace*{0.5cm}

\end{center}

\begin{abstract}
\noindent
We present a top-condensation model in which the $CP$ symmetry is spontaneously broken at the electroweak scale due to the condensation of two composite Higgs doublets.
In particular the $CP$-violating phase of the CKM matrix is generated.
A simpler model where only one quark family is included is also discussed.
In this case, for a general four-fermion interaction
($G_{tb}\neq 0$),
the particle spectrum is the one of the one Higgs doublet model.
\end{abstract}

\vskip 2.5cm


\vfill

\end{titlepage}


\section{Introduction}

Top-condensation models
\cite{NAMBU89,Miransky:1988xi,Miransky:1989ds,Bardeen:1989ds,Suzuki:1989nv,Cvetic:1997eb}\footnote{See also \cite{He:1998ie,Yue:2004jc,Larios:2004jx,Cao:2004wd,Yue:2005qk,Wang:2005ra} for more phenomenological aspects.} are interesting candidates for a realistic theory of electroweak symmetry breaking (EWSB).
They are a particular case of models of dynamical EWSB
\cite{Hill:2002ap},
where the electroweak (EW) gauge symmetry is broken due to the condensation of fermion-antifermion bilinears.
In top-condensation models the condensates are made of known fermions, mainly of the top quark.
In the minimal models no new particles are postulated.
The fundamental degrees of freedom are only fermions and gauge bosons.
No fundamental scalar fields such as the standard model (SM) Higgs field are present.
On the other hand the spectrum of the theory includes one or more composite Higgs particles.
They are fermion-antifermion bound states and play a similar role as the fundamental Higgs particles in the SM and its extensions.

Besides the well-established $SU(3)\times SU(2)_L \times U(1)_Y$ gauge interactions a new four-fermion interaction is postulated.
It provides the dynamics responsible of EWSB and
the generation of SM fermion masses at the same time.
Note that in general two different sectors are necessary for these purposes.
In technicolor, for example, technicolor gauge interactions trigger EWSB while extended technicolor interactions are required to generate fermion masses.
Due to the non-perturbative nature of the new interaction some approximation is needed.
Calculations at first order in the $1/N$ expansion, where $N=3$ is the number of colors are often made, obtaining a version of the Nambu-Jona-Lasinio (NJL) model \cite{Nambu:tp}.
Next-to-leading order calculation are also available
\cite{Cvetic:1994cc,Cvetic:1995qn}.

The four-fermion term postulated in top-condensation models is generally seen as an effective interaction.
At some high energy scale $\Lambda$ an asymptotically free gauge interaction becomes strongly-coupled.
For energies below $\Lambda$ the new interaction is effectively described by operators constructed with the fields corresponding to the light ($m_\text{particle}<\Lambda$) degrees of freedom of the theory.
At low energies the most important non-renormalizable operators are the ones having the lowest mass dimension.
Therefore dimension-six four-fermion operators are considered.
Normally only four-fermion operators made of (pseudo)scalar fermion bilinears are taken into account.
They are the ones leading to (pseudo)scalar composite fields.
Axial vector and vector fermion bilinears are ignored.
Note, however, that the distinction between (pseudo)scalar and (axial)vector bilinears is ambiguous due to Fierz identities.
The new interaction must violate the flavor symmetry,
i.e. must be non-universal, in order to generate the observed fermion mass pattern.
Topcolor models \cite{Hill:1991at} are examples of  a theory of this type.
In a second scenario the four-fermion interaction term acquires a more fundamental status.
It is assumed that
the SM with the Higgs sector being replaced by a general dimension-six four-fermion interaction is a (non-perturbatively) renormalizable theory \cite{Gies:2003dp}.
This is the case if one or more non-Gaussian ultraviolet stable fixed points are established  beyond the point-like approximation  \cite{Gies:2003dp}.
In this note the four-fermion interaction term is used as a starting point.
Therefore, besides the four-fermion and the SM gauge couplings,
the scale $\Lambda$ at which the whole Lagrangian is defined, is also a parameter of the theory.

For $\Lambda$ much bigger than the electroweak scale,
complementary to the NJL approach
a perturbative renormalization group analysis can be made
\cite{Bardeen:1989ds,Luty:1990bg,Froggatt:1990wa,
Froggatt:1992wt,Mahanta:1991jc,Cvetic:1993xg}.\footnote{In this case, however, the $1/N$ expansion is difficult to justify \cite{Cvetic:1994cc,Cvetic:1995qn}.}
This method, which incorporates the SM gauge interactions, provides reliable values for the top-quark and Higgs-boson masses.
In this approach the information that composite Higgs doublets appear at the scale $\Lambda$ is encoded in the compositeness condition.
To have a very high scale $\Lambda$ is, however, not very attractive because the  theory suffers from fine-tuning in exactly the same way as the SM.
Another important point is the one related to the distinction between fundamental and  composite Higgs particles.
If the compositeness scale $\Lambda$ is very high,
the composite nature of these particles cannot be directly verified by experiments in the near future.
A more interesting possibility is to have a scale $\Lambda$ not very much higher than the EW scale, $\Lambda\sim 5-10\; TeV$.
In this case no fine-tuning problem appears.
Besides, the generation of the scale  $\Lambda$ could be explained from dimensional transmutation.
This would solve or avoid the hierarchy problem.
A perturbative renormalization group analysis cannot be justified in this case.
Topcolor assisted technicolor \cite{Hill:1994hp},
and top-quark see-saw \cite{Dobrescu:1997nm,Chivukula:1998wd} are examples of  theories of this type.
In these theories the NJL approach is widely used.

In this paper, we present a top-condensation model in which together with the EW symmetry the discrete $CP$ symmetry is spontaneously broken.
A $CP$-conserving four-fermion interaction term including the three quark generations is considered.
Under certain conditions the $CP$ symmetry is spontaneously broken and the $CP$-violating phase of the Cabibbo-Kobayashi-Maskawa (CKM) matrix is generated.
Besides, realistic mixing angles and quark masses are obtained.
Leading order approximation in the $1/N$ expansion is adopted.
A more detailed treatment of the model is presented in \cite{Valenzuela:2005th}.

The first model with spontaneous $CP$ violation was considered by T.~D.~Lee in 1973 \cite{Lee:iz}, the same year that M.~Kobayashi and T.~Maskawa published their important paper \cite{Kobayashi:1973fv}.
In the Lee model (or its generalization for the case of 3 quark families) two fundamental Higgs fields are considered.
For certain values of the parameters of the model the $CP$-symmetric effective potential has a $CP$-non-symmetric minimum.
As a result the model has a number of $CP$-violating interactions, namely $W^\pm$ boson exchange (as in the SM), charged and (flavor violating and flavor conserving) neutral Higgs boson exchange.
Flavor changing neutral currents (FCNCs) are present in this model already at tree level.
This requires a mechanism for their suppression in order to avoid conflict with experimental data.

There are two potential problems in models of spontaneous $CP$-symmetry breaking at the EW scale.
The first is the one related with FCNCs.
It can be shown \cite{Gronau:1987xz,Branco:1979pv} that the requirements of spontaneous $CP$ symmetry breaking at the EW scale, absence of FCNCs at tree level\footnote{Here we refer to the general case without assuming any discrete symmetry.}, and a realistic CKM matrix cannot be simultaneously satisfied.
Therefore, models presenting spontaneous $CP$-symmetry breaking at the EW scale have tree-level non-diagonal Yukawa coupling matrices.
In order to have a realistic model a suppression mechanism is needed.
The second potential problem is a domain wall problem which arises in a cosmological context.
The effective potential has in general degenerate minima corresponding to complex conjugate vacuum expectation values (VEVs).
At the EW phase transition domains with different signs of the VEV phases are formed.
These domains are separated by walls with energy density much bigger than the closure energy of the universe
(after taking into account the effect of the universe expansion) \cite{Zeldovich:1974uw}.
If one considers this problem to be a serious one, some solution must be found in order that spontaneous $CP$ symmetry breaking at the EW scale is viable.

This paper is organized as follows:
we begin in section \ref{secc:minscheme} by presenting the minimal scheme in top-condensation models.
We continue, in section \ref{secc:1fam}, with a model having a four-fermion interaction term which include only quarks of the third family. Section \ref{secc:3fam} corresponds to the central part of this paper. Here we consider a model with a four-fermion interaction term including the three quark families.
In this model the $CP$ symmetry is broken by the vacuum.
In section \ref{secc:conclusions} the conclusions are given.
In appendix \ref{app:eff_potential} the calculation of the effective potential in the leading $1/N$ expansion is sketched.
In appendix \ref{app:auxfield_bases} convenient basis changes for the auxiliary fields are given.
In appendices \ref{app:two-point_1fam} and \ref{app:two-point_3fam} composite-field two-point functions are calculated for the models of sections \ref{secc:1fam} and \ref{secc:3fam}, respectively.
In appendix \ref{app:some_definitions} some definitions used in section \ref{secc:3fam} are given.

\section{Minimal Scheme}\label{secc:minscheme}

We start with the simplest Lagrangian leading to EWSB in the context of top-condensation models.
$CP$ is in this case a symmetry of the Lagrangian and of the vacuum.
In top-condensation models the SM Lagrangian without the Higgs sector is considered.
In its place a four-fermion interaction
term made of SM fermions is postulated:

\begin{equation}\label{min1}
\mathcal{L}= \sum_k \;\bar{\Psi}_k \:i \gamma^\mu D_\mu \,\Psi_k
    -\frac{1}{4} \sum_i \;(F_{\mu\nu}^{(i)\,a})^2
    +\mathcal{L}_{\text{4f}},
\end{equation}
where the first sum is over all left- and right-handed fermions of
the theory and the second contains the 3 Yang-Mills terms of the
SM-symmetry group, $SU(3)_c\times SU(2)_L\times U(1)_Y$. The
Lagrangian $\mathcal{L}$ is locally invariant under this symmetry
group. In the simplest model \cite{Bardeen:1989ds} the four-fermion
interaction term $\mathcal{L}_{\text{4f}}$, is given by

\begin{equation}\label{min2}
\mathcal{L}_{\text{4f}} = G_t\, (\bar{\psi}_L t_R)
(\bar{t}_R\psi_L),
\end{equation}
where $\psi_L=(t_L,\, b_L)^T$, and $t$ and $b$ are the top and
bottom quark fields. The $SU(3)_c$ and $SU(2)_L$ indices are
suppressed. A color-index contraction in each parenthesis and a
$SU(2)_L$-index contraction between $\bar{\psi}_L$ and $\psi_L$ are
understood.
The coupling constant $G_t$ has mass dimension $m^{-2}$.
The theory is defined at the scale $\Lambda$ with all heavier degrees of freedom integrated out.
All momentum integrals of the theory are
regularized using $\Lambda$ as a spherical cutoff.

In order to study the vacuum of the theory
it is very convenient to use
the auxiliary field formalism \cite{Gross:1974jv,Kugo:1976tq}, especially if one goes beyond the minimal scheme
as we do in the next sections.
The formalism is also useful for studying next-to-leading order corrections in the $1/N$ expansion
\cite{Cvetic:1994cc,Cvetic:1995qn,Cvetic:1996uq}.\footnote{To see the connection between the formalisms with and without auxiliary fields in the case of one auxiliary field, see \cite{Cvetic:1992nh}.
In this paper the effective potential is calculated diagrammatically.}
Let us introduce a scalar auxiliary field $H$ which possesses the same quantum numbers as the fermion bilinear $\bar{t}_R\psi_L$,
i.e., the quantum numbers of the Higgs doublet field in the SM.
The term $\mathcal{L}_{\text{4f}}$ is replaced by $\mathcal{L}_{\text{aux}}$:

\begin{equation}\label{}
\mathcal{L}_{\text{4f}} \longrightarrow
\mathcal{L}_{\text{aux}}=
-m_H^2 H^\dagger H
-g_t(\bar{\psi}_L t_R\: H + h.c.),
\end{equation}
with real mass parameter $m_H^2$ and Yukawa coupling constant $g_t$.
The Lagrangian $\mathcal{L}_{\text{aux}}$ depends on the auxiliary field $H$ only quadratically.
One recovers $\mathcal{L}_{\text{4f}}$, with $G_t=g_t^2/m_H^2$,
either integrating out the field $H$ from the generating functional in the path integral formulation, or imposing the constrains over the non-dynamical field $H$ (Euler-Lagrange equations).

The vacuum of the theory is obtained minimizing the effective potential related to the field $H$.
From appendix \ref{app:eff_potential} we have the effective potential in the $N\rightarrow\infty$ limit:

\begin{equation}\label{min3}
V_{\text{eff}}(v^{'})=
m_H^2 \frac{v^{'2}}{2}
-\frac{N}{8\pi^2}\int_0^{\Lambda^2} k^2\,dk^2\:
\log(k^2+g_t^2v^{'2}/2),
\end{equation}
with $H^\dagger H=v^{'2}/2$.
From the minimum condition which involve the first derivative of $V_{\text{eff}}$ with respect to $v^{'}$ one gets the gap equation

\begin{equation}\label{min4}
v^{'}\left(
m_H^2 -\frac{g_t^2 N}{8\pi^2}\int_0^{\Lambda^2}
          \frac{k^2\,dk^2}{k^2+g_t^2v^{'2}/2}
\right)=0 .
\end{equation}
This equation has in general two solutions.
A symmetrical one located at $v^{'}=0$,
and, if $G_t>G_{\text{crit}}=8\pi^2/N\Lambda^2$, a second one with $v^{'}\neq 0$ which breaks the EW symmetry.
Evaluating the second derivative of $V_{\text{eff}}(v^{'})$ at these two points it can be seen that the non-symmetrical solution corresponds to the minimum of the effective potential if $G_t>G_{\text{crit}}$.
In this case the corresponding value of $v^{'}$ (or equivalently of $m_t$) at the minimum is given by the solution of

\begin{equation}\label{min5}
G_t\left(
1 -\frac{m_t^2}{\Lambda^2}\log\Lambda^2/m_t^2
\right)   =G_{\text{crit}},
\end{equation}
with $m_t^2=g_t^2v^{'2}/2$.
Note that in order to fulfill the last equation having $m_t^2\ll\Lambda^2$,
fine-tuning of $G_t$ is needed.
Thus, if $G_t>G_{\text{crit}}$ the electroweak symmetry is spontaneously broken to the $U(1)$ electromagnetic one.
As in the SM, three Goldstone bosons appear and the Higgs mechanism provides the gauge bosons with masses.

The fourth degree of freedom $\phi$ of the auxiliary field $H$ describes a scalar top-antitop bound state.
The inverse propagator of the field $\phi$ is given by

\begin{equation}\label{min6}
i\Gamma_{\phi,\phi}(p^2)=\frac{i g_t^2 N}{16\pi^2}
(p^2-4m_t^2) \int_0^{\Lambda^2}
\frac{k^2\,dk^2}{[(p+k)^2+m_t^2](k^2+m_t^2)}.
\end{equation}
The propagator has a pole located at $p^2=(2m_t)^2$.
The model predicts in this approximation a scalar bound state with a mass equal to twice the top-quark mass\footnote{If one modifies $\mathcal{L}_{\text{4f}}$ in eq. (\ref{min1}) in order to provide all the 6 quarks with masses in such a way that $\mathcal{L}_{\text{4f}}$ can be rewritten with help of only one composite Higgs field (see below),
then the mass of the bound state remains almost unaffected $\approx 2m_t$.}.
The auxiliary field $H$ plays a similar role as the Higgs field in the SM.
For this reason we call it (composite) Higgs field in the following.

\section{Third-generation Case}\label{secc:1fam}

We generalize eq. (\ref{min2}) for the case in which the four chiral fields of the third family of quarks interact.
We consider the most general (dimension 6) gauge-invariant four-quark interaction term involving the four chiral fields that can be written as a sum of products of fermion bilinears with the quantum numbers and Lorentz structure of the SM Higgs boson

\begin{equation}\label{1fam1}
\mathcal{L}_{\text{4f}} =
G_t\,   (\bar{\psi}_L t_R) (\bar{t}_R\psi_L)+
G_b\,   (\bar{\psi}_L b_R) (\bar{b}_R\psi_L)+
\left[ G_{tb} \,\epsilon^{ab}
        (\bar{\psi}_L^a b_R)(\bar{\psi}_L^b t_R)
+ h.c. \right],
\end{equation}
where
$\epsilon^{ab}=
\begin{pmatrix}
0 & -1 \\
1 &  0
\end{pmatrix}$.\\
Due to the hermiticity of the Lagrangian $G_t$ and $G_b$ are real.
One can set $G_{tb}$ also real (or positive) by redefining one of the right-handed fermion fields.
In this way the interaction term $\mathcal{L}_{\text{4f}}$
possesses only real coupling constants
and the complete Lagrangian is invariant under a conventional $CP$
transformation\footnote{We ignore here the QCD $\theta$-term.}.

Repeating the procedure followed in the previous section we introduce $n$ spin-zero auxiliary fields $H^{(i)}$.
The term $\mathcal{L}_{\text{4f}}$ is replaced by

\begin{equation}\label{1fam2}
\mathcal{L}_{\text{aux}}=
- \sum_{i=1}^n m_{H_i}^2 H^{(i)\dagger}H^{(i)}+\mathcal{L}_{\text{Yukawa}},
\end{equation}
with
\begin{equation}\label{1fam3}
\mathcal{L}_{\text{Yukawa}} = -\sum_{i=1}^n\;
(\;g_t^{(i)}\:\bar{\psi}_L t_R\:H^{(i)}+
   g_b^{(i)}\:\epsilon^{ab}\:\bar{\psi}_L^a b_R\:H^{b(i)*}
\;+\; h.c.\;),
\end{equation}
where $g_t^{(i)}$ and $g_b^{(i)}$ are the real Yukawa coupling constants
and $m_{H_i}^2$ are mass parameters associated with the auxiliary fields.
The relations between the coupling constants in the two formulations of the model are given by

\begin{equation}\label{1fam4}
\begin{split}
G_t   &=\sum_{r=1}^n\;\frac{g_t^{(r)}g_t^{(r)}}{m_{H_r}^2},\\
G_b   &=\sum_{r=1}^n\;\frac{g_b^{(r)}g_b^{(r)}}{m_{H_r}^2},\\
G_{tb}&=\sum_{r=1}^n\;\frac{g_t^{(r)}g_b^{(r)}}{m_{H_r}^2}.
\end{split}
\end{equation}
In order to parameterize the space of couplings $G$,
it is enough to consider $n=2$ .
We restrict ourselves to $n=2$  in the following.

From appendix \ref{app:eff_potential} we obtain the effective potential in the $N\rightarrow\infty$ limit:

\begin{equation}\label{1fam5}
V_{\text{eff}}(H^{(1)},H^{(2)})=
\sum_{i=1,2} m_{H_i}^2 H^{(i)\dagger}H^{(i)}
-\frac{N}{8\pi^2}\int_0^{\Lambda^2} k^2\,dk^2\:
\log\det(k^2+A),
\end{equation}
with the $2\times 2$ matrix $A$ given by

\begin{equation}\label{1fam6}
A=
\begin{pmatrix}
 g_t^{(i)}g_t^{(j)} H^{(i)\dagger}H^{(j)} &
-g_t^{(i)}g_b^{(j)} \epsilon^{ab} H^{a(i)*}H^{b(j)*} \\
-g_t^{(i)}g_b^{(j)} \epsilon^{ab} H^{a(i)} H^{b(j)} &
 g_b^{(i)}g_b^{(j)} H^{(i)\dagger}H^{(j)}
\end{pmatrix},
\end{equation}
where summation over the indices $i$ and $j$ is understood.

The ground state of the theory is found by minimizing the effective potential
with respect to the auxiliary fields $H^{(1)}$ and $H^{(2)}$.
Due to the gauge invariance it is possible to gauge any field configuration into the following form:

\begin{equation}\label{1fam7}
H^{(1)}=\bigg(\begin{matrix}
              \frac{v^{'}}{\sqrt{2}}\\0
              \end{matrix}\bigg),\qquad\qquad
H^{(2)}=\bigg(\begin{matrix}
              \frac{w^{'}e^{i\eta^{'}}}{\sqrt{2}}\\z{'}
              \end{matrix}\bigg),
\end{equation}
with $v^{'}$, $w^{'}$ , $z^{'}\geq 0$.
In the following $v^{'}$, $w^{'}$, $z^{'}$, $\eta^{'}$ denote the classical fields
and the corresponding non-primed symbols denote their VEVs,

\begin{equation}\label{1fam8}
<H^{(1)}>\,=\bigg(\begin{matrix}
              \frac{v}{\sqrt{2}}\\0
              \end{matrix}\bigg),\qquad\qquad
<H^{(2)}>\,=\bigg(\begin{matrix}
              \frac{w\;e^{i\eta}}{\sqrt{2}}\\z
              \end{matrix}\bigg).
\end{equation}
In order to preserve the electromagnetic $U(1)$ symmetry, the VEV $z$ must be zero.
Besides, if the phase $\eta$ is not a trivial one,
the $CP$ symmetry could be spontaneously broken.
This happens if, once non-trivial quark masses are generated, the coupling constant $G_{tb}$ in the fermion mass basis is complex.

Next we minimize the effective potential with respect to the variables $v^{'}$, $w^{'}$, $\eta^{'}$, and $z^{'}$.
We shall restrict ourselves to the parameter subspace with $z^{'}=0$
and search for local minima in this region.
It is possible to show \cite{Harada:1990wg} that for $z^{'}\neq 0$ there is no local minimum
(at least for $m_t\neq m_b$).
The following conditions are sufficient in order to have a local minimum
at a point with $z^{'}= 0$:
\begin{equation}\label{1fam10}
\begin{split}
&\text{a)}\; \frac{\partial V_{\text{eff}}}{\partial\theta} = 0\;,
         \text{ for } \theta=v^{'},w^{'},\eta^{'}, \\
&\text{b)}\; \frac{\partial V_{\text{eff}}}{\partial z^{'2}} > 0,
        \quad\text{(it is convenient to use  $z^{'2}$ instead of $z^{'}$),}\\
&\text{c)}\; \text{The $3\times 3$ Hessian matrix associated with the variables} \\
&\quad \text{$v^{'}$, $w^{'}$ and $\eta^{'}$ is positive definite.}
\end{split}
\end{equation}

The conditions a) evaluated at the point
$v^{'}=v$, $w^{'}=w$, $\eta^{'}=\eta$, and $z^{'}=0$
are given by

\begin{equation}\label{1fam11}
v \bigg[ m_{H_1}^2-
\frac{N\Lambda^2}{8\pi^2}\sum_{q=t,b}
\bigg(1-\frac{m_q^2}{\Lambda^2}\log\Lambda^2/m_q^2 \bigg)
\Big( (g_q^{(1)})^2+g_q^{(1)}g_q^{(2)} \frac{w}{v}\cos\eta \Big) \bigg]
=0,
\end{equation}

\begin{equation}\label{1fam12}
w \bigg[ m_{H_2}^2-
\frac{N\Lambda^2}{8\pi^2}\sum_{q=t,b}
\bigg(1-\frac{m_q^2}{\Lambda^2}\log\Lambda^2/m_q^2 \bigg)
\Big( (g_q^{(2)})^2+g_q^{(1)}g_q^{(2)} \frac{v}{w}\cos\eta \Big) \bigg]
=0,
\end{equation}

\begin{equation}\label{1fam13}
v w \sum_{q=t,b}
\bigg(1-\frac{m_q^2}{\Lambda^2}\log\Lambda^2/m_q^2 \bigg)
g_q^{(1)}g_q^{(2)} \sin\eta
=0,
\end{equation}
where
\begin{equation}\label{1fam14}
m_q=\bigg| g_q^{(1)}\frac{v}{\sqrt{2}}+g_q^{(2)}
\frac{w\, e^{i\eta}}{\sqrt{2}}
\bigg|.
\end{equation}
The first derivative of the effective potential with respect to $z^{'2}$ is given by

\begin{equation}\label{1fam15}
\begin{split}
\frac{\partial V_{\text{eff}}}{\partial z^{'2}}=
& \;m_{H_2}^2-
\frac{N\Lambda^2}{8\pi^2}\sum_{q=t,b}
\bigg(1-\frac{m_q^2}{\Lambda^2} \log\Lambda^2/m_q^2 \bigg)
(g_q^{(2)})^2 \\
&+\frac{N}{8\pi^2}
\int_0^{\Lambda^2}
\frac{k^2 \, dk^2}{(k^2+m_t^2)(k^2+m_b^2)}\frac{v^2}{2}
(g_t^{(1)}g_b^{(2)}-g_t^{(2)}g_b^{(1)})^2.
\end{split}
\end{equation}
Using eq. (\ref{1fam12}), with $w\neq 0$, the last expression can be written as

\begin{equation}\label{1fam16}
\begin{split}
\frac{\partial V_{\text{eff}}}{\partial z^{'2}}=
& \;
\frac{N\Lambda^2}{8\pi^2}\sum_{q=t,b}
\bigg(1-\frac{m_q^2}{\Lambda^2}\log\Lambda^2/m_q^2 \bigg)
g_q^{(1)}g_q^{(2)} \frac{v}{w}\cos\eta \\
&+\frac{N}{8\pi^2}
\int_0^{\Lambda^2}
\frac{k^2 \, dk^2}{(k^2+m_t^2)(k^2+m_b^2)}\frac{v^2}{2}
(g_t^{(1)}g_b^{(2)}-g_t^{(2)}g_b^{(1)})^2.
\end{split}
\end{equation}

Because we are studing the possibility of having spontaneous $CP$ symmetry breaking,
the interesting case is when both auxiliary field doublets condense ($v,\, w\neq 0$) and $G_{tb}\neq 0$.\footnote{We briefly discuss the case $G_{tb}= 0$ below.}
We look for a local minimum of this form.
However,  using $G_{tb}\neq 0$ we immediately note that eq. (\ref{1fam13}) can only be fulfilled if

\begin{equation}\label{h24a}
\sin\eta=0.
\end{equation}
Thus, Higgs VEVs are all real and $CP$ is a symmetry of the theory.

The first derivative conditions eqs. (\ref{1fam11}) and  (\ref{1fam12}) determine the top and bottom quark masses.
These equations can be fulfilled by choosing the adequate values of the mass parameters $m_{H_1}^2$ and $m_{H_2}^2$.
In a similar way as for the minimal scheme, these conditions are fine-tuned if $m_q\ll \Lambda$.

Considering only terms of order $\Lambda^2$ in eq. (\ref{1fam16}), condition b) is equivalent to

\begin{equation}\label{1fam16b}
(g_t^{(1)}g_t^{(2)}+g_b^{(1)}g_b^{(2)})\cos\eta >0.
\end{equation}
For given couplings $g_q^{(i)}$, the last inequality can be fulfilled by choosing the sign of $\cos\eta$  properly.

Finally we check condition c).
The Hessian of the effective potential with respect to $v^{'}$, $w^{'}$, and $\eta^{'}$ is given by

\begin{equation}\label{1fam17}
\begin{split}
\frac{\partial^2 V_{\text{eff}}}{\partial\theta_a\partial\theta_b}&=
\frac{N\Lambda^2}{8\pi^2} \sum_{q=t,b}
\bigg(1-\frac{m_q^2}{\Lambda^2}\log\bigg(\frac{\Lambda^2}{m_q^2}+1\bigg)\bigg)
g_q^{(1)}g_q^{(2)} \cos\eta
\begin{pmatrix}
w/v & -1  & 0  \\
-1  & v/w & 0  \\
0   &  0  & v w
\end{pmatrix}    \\
&\quad +\frac{N}{8\pi^2} \sum_{q=t,b}
\:\int_0^{\Lambda^2} 2 m_q^2 \,\frac{dx\, x}{(x+m_q^2)^2}
\begin{pmatrix}
(g_q^{(1)})^2               & g_q^{(1)}g_q^{(2)} \cos\eta & 0 \\
g_q^{(1)}g_q^{(2)} \cos\eta & (g_q^{(2)})^2               & 0 \\
0 & 0 & 0
\end{pmatrix} ,
\end{split}
\end{equation}
with $\theta_a=v^{'},w^{'},\eta^{'}$.
To obtain eq. (\ref{1fam17}) the first derivative conditions were used.
Considering condition b) which determines the sign of $\cos\eta=\pm 1$ one can see that the last matrix has 3 positive eigenvalues.

In summary, we considered the model defined by (\ref{1fam1}) with  $G_{tb}\neq 0$ when both auxiliary fields condense breaking the EW symmetry but respecting the $U(1)$ electromagnetic symmetry.
This occurs if eqs. (\ref{1fam11}) and (\ref{1fam12}) are fulfilled.
For $m_q\ll \Lambda$ these two equations are fine-tuned.
The sign of $\cos\eta=\pm 1$ is determined by condition (\ref{1fam16b}).
Furthermore spontaneous $CP$ symmetry breaking does not occur.
The spectrum of the model is calculated in appendix \ref{app:two-point_1fam}.
We found that only 4 of the 8 degrees of freedom related to the two auxiliary Higgs doublets are relevant at low energies.
Three of them are Goldstone bosons and the fourth is a physical Higgs particle with mass $\approx 2m_t$.
At energies much lower than $\Lambda$ this model cannot be distinguished from the one which has only one auxiliary Higgs doublet from the beginning
(in this case the condition $G_t G_b = G_{tb}^2$ must hold, see eq. (\ref{1fam4})).

The case of having $G_{tb}= 0$ is qualitatively very different because the Lagrangian possesses a Peccei-Quinn $U(1)$ symmetry.
In this case it is possible to introduce two auxiliary fields in such a way that one couples only to the field $t_R$ and the other only to the field $b_R$ (2 Higgs doublet (2HD) model type II).
If $G_t, G_b > G_{\text{crit}}$ both auxiliary fields condense and
one obtains a particle spectrum as follows:
In the neutral sector two scalars with masses equal to $2m_t$ and $2m_b$,
a Goldstone boson, and an axion appear.\footnote{An axion at the electroweak scale is experimentally ruled out.}
In the charged sector one obtains a charged Goldstone boson and a charged particle with mass $\approx \sqrt{2(m_t^2+m_b^2)}$.

We see that both, for the case of having $G_{tb}= 0$ and for $G_{tb}\neq 0$ (at least when both auxiliary fields condense) the number of parameters at low energies is two.
Once one fixes them, e.g. the quark masses, the model is completely specified.
For the case $G_{tb}\neq 0$ one would like to understand why the number of parameters is only two and not three.
The reason is that in order to satisfy eqs. (\ref{1fam11}) and (\ref{1fam12})
simultaneously (with $G_{tb}\neq 0$), the following relation must hold:

\begin{equation}\label{1fam18}
\frac{1}{G_\text{crit}}  (G_t+G_b)-
\frac{1}{G_\text{crit}^2}(G_t G_b - G_{tb}^2)
=1+\mathcal{O}(m_q^2/\Lambda^2).
\end{equation}
For a given $G_{tb}$ ($\neq 0$), the values of $G_t$ and $G_b$ which fulfill the last equation
describe a hyperbola.

\section{Spontaneous $CP$ symmetry breaking}\label{secc:3fam}

In order to generate the $CP$-violating phase of the CKM matrix and thus reproduce the observed mechanism of $CP$ symmetry violation, quarks of the three generations must be included in the four-fermion interaction.
We are interested in the situation in which the $CP$ symmetry is spontaneously broken by the vacuum.
Therefore our starting point is a $CP$-invariant Lagrangian with the four-fermion interaction given by a generalization of eq. (\ref{1fam1}):

\begin{equation}\label{3f1}
\mathcal{L}_{\text{4f}} =
G_{ijkl}\,(\bar{\psi}_{iL} u_{jR}) (\bar{u}_{lR}\psi_{kL})+
G_{ijkl}^{'}\,(\bar{\psi}_{iL} d_{jR}) (\bar{d}_{lR}\psi_{kL})+
[G_{ijkl}^{''}\,\epsilon^{ab}(\bar{\psi}_{iL}^a d_{jR})
    (\bar{\psi}_{kL}^b u_{lR})+ h.c. ],
\end{equation}
where the coupling constants $G$ and the quark fields $u_R$, $d_R$, $\psi_L=(u_L,d_L)^T$ have indices $i$, $j$, $k$, $l$,
which go from the first to the third quark generation.
The Lagrangian $\mathcal{L}_{\text{4f}}$ includes four-fermion terms which mix quark fields of different quark families.

Because we demand the Lagrangian to be $CP$-invariant,
all the coupling constants $G$ in $\mathcal{L}_{\text{4f}}$ are considered to be real.
In this case the number of parameters of eq. (\ref{3f1}) is 171.\footnote{After requiring hermiticity of the Lagrangian there are 45 independent couplings $G$, 81  couplings $G^{'}$, and further 45 couplings $G^{''}$.}
Where we do not count the cutoff $\Lambda$ which is also a parameter of the model.

We now rewrite the interaction term, eq. (\ref{3f1}), in terms of auxiliary fields which have the quantum numbers of the SM Higgs doublet field as we did in the previous sections.
We do not consider completely arbitrary couplings $G$.
For simplicity we restrict ourselves to the subset for which the theory can be described by means of only two auxiliary fields,
$H^{(1)}$ and  $H^{(2)}$.
In terms of these auxiliary fields the four-fermion term $\mathcal{L}_{\text{4f}}$ is replaced by

\begin{equation}\label{3f2}
\mathcal{L}_{\text{4f}} \longrightarrow
\mathcal{L}_{\text{aux}}=
- \sum_{i=1}^2 m_{H_i}^2 H^{(i)\dagger}H^{(i)}+\mathcal{L}_{\text{Yukawa}},
\end{equation}
where
\begin{equation}\label{3f3}
\mathcal{L}_{\text{Yukawa}} = -\sum_{i=1}^2\;
(\;g_{kl}^{(i)}\:\bar{\psi}_{kL}u_{lR}\:H^{(i)}+
 h_{kl}^{(i)}\:\epsilon^{ab}\:\bar{\psi}_{kL}^a d_{lR}\:H^{b(i)*}
\;+\; h.c.\;),
\end{equation}
with $g^{(i)}$ and $h^{(i)}$ $3\times 3$ real matrices.
The relations between the real coupling constants in the two formulations of the model are given by

\begin{equation}\label{int6}\begin{split}
G_{ijkl}&=\sum_{r=1}^n\; \frac{g_{ij}^{(r)}g_{kl}^{(r)}}{m_{H_r}^2},\\
G_{ijkl}^{'}&=\sum_{r=1}^n\;\frac{h_{ij}^{(r)}h_{kl}^{(r)}}{m_{H_r}^2},\\
G_{ijkl}^{''}&=\sum_{r=1}^n\;\frac{h_{ij}^{(r)}g_{kl}^{(r)}}{m_{H_r}^2}.
\end{split}\end{equation}

In consequence we restrict ourselves to a model with 36,
essentially the four $3\times 3$ Yukawa matrices, parameters
(plus the cutoff scale $\Lambda$).

We are now confronted with the following problem.
We want to find the values of the parameters of the model
such that the vacuum of the theory breaks the EW symmetry in the observed way
(spontaneous $CP$ symmetry violation keeping the electromagnetic $U(1)$ symmetry unbroken).
Besides, the generated CKM matrix and the quark masses must correspond to their measured values.
However, the relation between the 36 parameters
of our model and the quantities to be reproduced is rather complicated.
In order to find an analytical solution we introduce a self-consistent approach to the problem.

First, we assume that the minimum of the effective potential
(denoted by non-primed symbols) is given by a field configuration with non-trivial values of $v$, $w$, and $\eta$ ($\neq 0,\pi$) and with $z=0$:

\begin{equation}\label{}
<H^{(1)}>\;=\bigg(\begin{matrix}
              \frac{v}{\sqrt{2}}\\0
              \end{matrix}\bigg),\qquad\qquad
<H^{(2)}>\;=\bigg(\begin{matrix}
              \frac{w\, e^{i\eta}}{\sqrt{2}}\\0
              \end{matrix}\bigg).
\end{equation}
VEVs of this form are necessary in order to have a theory with spontaneous $CP$ symmetry breaking and unbroken $U(1)_{em}$ symmetry.

Inserting the Higgs VEVs in the Yukawa interactions, eq. (\ref{3f3}), one gets the quark mass term:

\begin{equation}\label{3f5}
\mathcal{L}_{\text{m}} = -\sum_{i=1}^2\;
(\;g_{kl}^{(i)}\:\bar{u}_{kL} u_{lR}\:<\phi^{0(i)}> +
 \;h_{kl}^{(i)}\:\bar{d}_{kL} d_{lR}\:<\phi^{0(i)}>^*
\;+\; h.c.\;),
\end{equation}
with $H^{(i)}=(\phi^{0(i)},\phi^{-(i)})^T$.
The quark mass matrices for the up- and down-type quarks in the last equation are in general non-diagonal.
In order to diagonalize them we perform the following chiral rotations:

\begin{equation}\label{3f6}
\begin{matrix}
u_{iR} = W_{ij}^u \;u_{jR}^{'}, & & u_{iL} = U_{ij}^u \;u_{jL}^{'},\\
&&\\
d_{iR} = W_{ij}^d \;d_{jR}^{'}, & & d_{iL} = U_{ij}^d \;d_{jL}^{'},
\end{matrix}
\end{equation}
where the primed fields denote the fermion fields in the mass basis and
$U^u$, $U^d$, $W^u$, $W^d$ are basis transformation matrices.
The CKM matrix is given by $V_{CKM}=U^{u\dagger}U^d$.
In the new basis the quark mass matrices, which are now diagonal and real are given by

\begin{equation}\label{3f7}
\begin{split}
M_u  &= \lambda_u^{(1)}\;\frac{v}{\sqrt{2}}+
         \lambda_u^{(2)}\;\frac{w\, e^{i\eta}}{\sqrt{2}}, \\
M_d  &= \lambda_d^{(1)}\;\frac{v}{\sqrt{2}}+
         \lambda_d^{(2)}\;\frac{w\, e^{i\eta}}{\sqrt{2}},
\end{split}\end{equation}
where the Yukawa couplings in the mass basis $\lambda_u^{(i)}$, $\lambda_d^{(i)}$ are defined by

\begin{equation}\label{3f8}
\begin{split}
g^{(i)} &\equiv U^u\;\lambda_u^{(i)}      \; W^{u\dagger},
                \quad\text{for }i=1,2, \\
h^{(i)} &\equiv U^d\;\lambda_d^{(i)\dagger}\;W^{d\dagger},
                \quad\text{for }i=1,2.
\end{split}\end{equation}
We emphasize that the Higgs VEVs in eq. (\ref{3f7}) are still not determined.

Using the last definitions, the relations between the composite fields $H^{(i)}$ and their constituent quark fields are given by:

\begin{equation}\label{int11b}
\begin{split}
\phi^{0(i)}&=-\frac{1}{m_{H_i}^2} \;\Big(
\bar{u}_R^{'}\:\lambda_u^{(i)\dagger}\:u_L^{'} +
\bar{d}_L^{'}\:\lambda_d^{(i)\dagger}\:d_R^{'}  \Big),  \\
\phi^{-(i)}&=-\frac{1}{m_{H_i}^2} \;\Big(
\bar{u}_R^{'}\:\lambda_u^{(i)\dagger}\:V_{CKM}\:d_L^{'} -
\bar{u}_L^{'}\:V_{CKM}\:\lambda_d^{(i)\dagger}\:d_R^{'}  \Big),
\end{split}
\end{equation}
for $i=1,2$.
Besides, the interaction term $\mathcal{L}_{\text{Yukawa}}$ is given in this basis by

\begin{equation}\label{int11}
\begin{split}
\mathcal{L}_{\text{Yukawa}} = -\sum_{i=1}^2\;\Big(\:
& \bar{u}_L^{'}\,\lambda_u^{(i)}\, u_R^{'}\:\phi^{0(i)}
   -\bar{u}_L^{'}V_{CKM}\,\lambda_d^{(i)\dagger}\, d_R^{'}\:\phi^{+(i)}  \\
&+\:\bar{d}_L^{'}V_{CKM}^\dagger\,\lambda_u^{(i)}\, u_R^{'}\:\phi^{-(i)}
   +\bar{d}_L^{'}\,\lambda_d^{(i)\dagger}\, d_R^{'}\:\phi^{0(i)*}
\;+\; h.c.\;\Big),
\end{split}
\end{equation}
with $\phi^{+(i)}\equiv\phi^{-(i)*}$.

Combining eqs. (\ref{3f7}) and (\ref{3f8}) and using the fact that the matrices
$g^{(i)}$ and  $h^{(i)}$ are real, it is possible to write the Yukawa
couplings in the weak basis, $g^{(i)}$, $h^{(i)}$, as a function of the Higgs VEVs, the quark masses, and the basis transformation matrices:

\begin{equation}\label{2hd5}
\begin{split}
g^{(1)} &= \frac{\sqrt{2}}{v}\;\big[\mathcal{R}e(U^u M_u W^{u\dagger})
                  -\cot\eta   \; \mathcal{I}m(U^u M_u W^{u\dagger})\big],\\
g^{(2)} &= \frac{\sqrt{2}}{w\,\sin\eta} \;
                                  \mathcal{I}m(U^u M_u W^{u\dagger}),\\
h^{(1)} &= \frac{\sqrt{2}}{v}\;\big[\mathcal{R}e(U^d M_d W^{d\dagger})
                  +\cot\eta   \; \mathcal{I}m(U^d M_d W^{d\dagger})\big],\\
h^{(2)} &= -\frac{\sqrt{2}}{w\,\sin\eta} \;
                                  \mathcal{I}m(U^d M_d W^{d\dagger}).\\
\end{split}\end{equation}

We have in this way transformed the original problem
into one which can be solved in a self-consistent way:
We must find values of $v$, $w$, and $\eta$,
to which we associate the Yukawa couplings given in Eqs. (\ref{2hd5}),
such that the resulting effective potential has its minimum at the same values $v$, $w$ and $\eta$ (besides $z=0$).
For this purpose we can vary the basis transformation
matrices $U^u$, $U^d$, $W^u$ and $W^d$.
These are arbitrary unitary matrices which must obey the condition
$U^{u\dagger}U^d=V_{CKM}$.

The effective potential related to the four-fermion interaction (\ref{3f1}) is given in  appendix \ref{app:eff_potential}.
A detailed analysis of the local minimum conditions given in (\ref{1fam10}) is presented in \cite{Valenzuela:2005th}.
The conditions obtained in order to have a model presenting spontaneous $CP$ symmetry violation, realistic quark masses, and the observed CKM matrix are:

\noindent i) The first derivative conditions associated with the variables $v^{'}$ and $w^{'}$ which are given by

\begin{equation}\label{2hd16}
\begin{split}
m_{H_1}^2&=\frac{N\Lambda^2}{8\pi^2}\sum_{i=1}^6
\bigg(1-\frac{m_i^2}{\Lambda^2}\log\bigg(\frac{\Lambda^2}{m_i^2}+1\bigg)\bigg)
\Big(\Sigma^{(1)}_{ii} + \frac{w^{'}}{v^{'}}\:
       \mathcal{R}e(\Sigma^{(0)}_{ii}e^{i\eta^{'}})\Big)\Big|_{(v,w,\eta,0)},\\
m_{H_2}^2&=\frac{N\Lambda^2}{8\pi^2}\sum_{i=1}^6
\bigg(1-\frac{m_i^2}{\Lambda^2}\log\bigg(\frac{\Lambda^2}{m_i^2}+1\bigg)\bigg)
\Big(\Sigma^{(2)}_{ii} + \frac{v^{'}}{w^{'}}\:
       \mathcal{R}e(\Sigma^{(0)}_{ii}e^{i\eta^{'}})\Big)\Big|_{(v,w,\eta,0)},\\
\end{split}
\end{equation}
where the $6\times 6$ matrices $\Sigma^{(i)}$ are given in appendix \ref{app:some_definitions}.
These conditions can easily be fulfilled by choosing suitable mass parameters $m_{H_i}^2$.
For $m_q\ll\Lambda$ both conditions require fine-tuning.

\noindent  ii) Complex Higgs VEVs are obtained only if the factor $c$, defined in appendix \ref{app:some_definitions}, is different from zero.
In this case, the first derivative condition associated with the variable  $\eta^{'}$ leads to:

\begin{equation}\label{2hd32}
\cot\eta= -\frac{a}{b+c},
\end{equation}
where the real factors $a$, $b$, and $c$ are given in appendix \ref{app:some_definitions}.
It can be seen that for $c\neq 0$ this first derivative condition also requires from fine-tuning if $m_q\ll \Lambda$.

\noindent  iii) The alignment condition
$\frac{\partial V_{eff}}{\partial z^{'2}}>0$ takes the following form:

\begin{equation}\label{2hd33}
-\frac{2\, c}{w^2 \sin^2\eta}
+\sum_{\begin{matrix}\scriptstyle i=u,c,t \\
                     \scriptstyle j=d,s,b   \end{matrix}}
\frac{|T_{ij}|^2}{\Lambda^2}
I(m_i^2,m_j^2) > 0,
\end{equation}
with the $3\times 3$  matrix $T$ and $I(m_i^2,m_j^2)$ given in appendix \ref{app:some_definitions}.

\noindent  iv) The Hessian matrix associated with the variables $v^{'}$, $w^{'}$, and $\eta^{'}$ must be positive definite.
The Hessian matrix is given by

\begin{equation}\label{3f25}
\begin{split}
\frac{\partial^2 V_{\text{eff}}}
     {\partial\theta_a\partial\theta_b}\Big|_{(v,w,\eta,0)}
&=
\:\:\frac{cN\Lambda^2}{4\pi^2\sin^2\eta}
\begin{pmatrix}
-1/v^2 &  1/vw  & 0 \\
1/vw   & -1/w^2 & 0 \\
0      &   0    & -1   \end{pmatrix}_{ab}  \\
&\quad+
\frac{N}{8\pi^2}\:\sum_{i,j=1}^6\:I(m_i^2,m_j^2)\:
\frac{\partial A_{ij}}{\partial\theta_a}
\frac{\partial A_{ji}}{\partial\theta_b}\Big|_{(v,w,\eta,0)},
\end{split}
\end{equation}
with $\theta_a,\theta_b=v^{'},w^{'}$, and $\eta^{'}$.
The first derivatives of the $6\times 6$ matrix $A$ are given in appendix \ref{app:some_definitions}.
There are no quadratically divergent terms in eq. (\ref{3f25}) ($c\propto 1/\Lambda^2$).
They cancel after imposing the fine-tuned first derivative conditions.

Once one has a set of parameters $g^{(i)}$, $h^{(i)}$, $m^2_{H_i}$ which fulfills the previous conditions, the corresponding couplings $G$ are obtained using eqs. (\ref{1fam4}).
For a detailed analysis of these conditions using a quark mass expansion see \cite{Valenzuela:2005th}.

The composite-field two-point functions, which allow us to find the composite Higgs masses, are calculated for the charged and neutral sectors in appendix \ref{app:two-point_3fam}.
Beside the three Goldstone bosons there are three neutral and one charged composite Higgs bosons.
In this approximation we find that one neutral Higgs mass is $\sim 2m_t$ and that the rest of the Higgs masses are much smaller.
We think this is a consequence of the crude approximation we adopted and not a property of the theory.
For further comments see our conclusions.

\section{Conclusions}\label{secc:conclusions}

In this paper we considered a top-condensation model with spontaneous $CP$ symmetry breaking.
We started with a $CP$ invariant four-fermion interaction defined at some scale $\Lambda$ which involve quarks of the three generations.
We restricted ourselves to the case in which the four-fermion interaction can be rewritten with help of two auxiliary fields.
The minimum of the effective potential is obtained in a self-consistent manner.
This leads to the relevant conditions on the parameters of the model in order to obtain a realistic CKM matrix, including the $CP$-violating phase, and the quark masses.
The following conditions are found:
i) first derivative conditions related to the variables $v^{'}$ and $w^{'}$ (eqs. (\ref{2hd16})),
ii) the condition for spontaneous $CP$ symmetry breaking $c\neq 0$,
iii) the preservation of the electromagnetic symmetry condition, and
iv) the second derivative conditions.
A more detailed treatment of these conditions including an analysis using a quark mass expansion can be found in \cite{Valenzuela:2005th}.

The spectrum of the theory corresponds to the one of the 2HD extension of the SM.
The Higgs particles are quark-antiquark bound states.
The related composite fields are given (at the scale $\Lambda$) by eq. (\ref{int11b}).
Thus, due to the hierarchy of the Yukawa couplings the composite Higgs particles are made mainly of the quarks of the third generation having their wave functions only a small light-quark component. In our approximation only one (neutral) composite Higgs has an acceptable mass value ($\approx 2m_t$).
The other  two neutral and the charged composite Higgs are unacceptably light.\footnote{Theories like topcolor assisted technicolor provide a mechanism for increasing the value of the composite Higgs boson masses.}
We think, this could change beyond the $1/N$ expansion used here.
For example for $\Lambda\gg\Lambda_{EW}$ a perturbative renormalization group approach
where the SM gauge interactions are taken into account, leads to much bigger composite Higgs masses \cite{Luty:1990bg,Froggatt:1990wa,
Froggatt:1992wt,Mahanta:1991jc,Cvetic:1993xg}.

Spontaneous $CP$ symmetry breaking at the EW scale presents two potential problems, namely, tree level FCNCs \cite{Gronau:1987xz,Branco:1979pv} and a domain wall problem
\cite{Zeldovich:1974uw}.
We just comment (in relation to the first potential problem)
that in the context of top-condensation models it is not understood why FCNCs are suppressed (if we simultaneously generate quark mixing angles)
independently of having spontaneous or explicit $CP$ symmetry violation.

We also treated a simpler top-condensation model where only quark fields belonging to the third generation are included in the four-fermion interaction (see eq. (\ref{min2})).
This Lagrangian is used as an intermediated step in theories like topcolor assisted technicolor \cite{Hill:1994hp} or top-quark see-saw \cite{Dobrescu:1997nm,Chivukula:1998wd}.
We found that in the general case
(without an extra Peccei-Quinn symmetry) the spectrum of the theory corresponds to the one of the SM with the composite neutral Higgs particle having a mass $\sim 2m_t$.
Further modes associated with masses of order $\Lambda$ are present. $\Lambda$ is, however, the upper limit of the validity range of the model.
For this reason these modes cannot be interpreted as particles.
In this simpler model spontaneous $CP$ symmetry breaking does not occur.



\section*{Acknowledgements}

I would like to thank Werner Bernreuther for useful discussions and support during all stages of this work.\\
This work was supported by the Deutschen Akademischen Austauschdienst (DAAD).

\begin{appendix}

\section{The Effective Potential}\label{app:eff_potential}

We present in this appendix the effective potential for $n$ scalar auxiliary fields $H^{(i)}$ coupled to 3 quark generations.
We consider the leading order contributions in the $1/N$ expansion,
i.e. in the $N\longrightarrow \infty$ limit keeping $G N$ fixed,
which is equivalent to the fermionic determinant approximation.
The Lagrangian is given by

\begin{equation}\label{A1}
\mathcal{L}=\mathcal{L}_{\text{kin}}
- \sum_{i=1}^n m_{H_i}^2 H^{(i)\dagger}H^{(i)}+\mathcal{L}_{\text{Yukawa}},
\end{equation}
where $\mathcal{L}_{\text{kin}}$ contains the quark kinetic terms and
\begin{equation}\label{A2}
\mathcal{L}_{\text{Yukawa}} = -\sum_{i=1}^n\;
(\;g_{kl}^{(i)}\:\bar{\psi}_{kL}u_{lR}\:H^{(i)}+
 h_{kl}^{(i)}\:\epsilon^{ab}\:\bar{\psi}_{kL}^a d_{lR}\:H^{b(i)*}
\;+\; h.c.\;),
\end{equation}
with complex parameters $g^{(i)}_{kl}$, $h^{(i)}_{kl}$,
and real mass parameters $m_{H_i}^2$.
No kinetic term for the auxiliary fields $H^{(i)}$ and no quartic term of the form $(H^\dagger H)^2$ are present.
The effective potential is given by

\begin{equation}\label{A3}
V_{\text{eff}}(\{H^{(i)}\})=m_{H_i}^2 H^{(i)\dagger}H^{(i)}
-i\int\frac{d^4k}{(2\pi)^4} \;
\log\det(\mathcal{D}^{-1}\{H^{(i)},k\}),
\end{equation}
where $\mathcal{D}^{-1}$ is the fermionic propagator in momentum space,
and is a function of the scalar fields $H^{(i)}$ and the momentum $k$.
After calculating the fermionic determinant one obtains

\begin{equation}\label{A4}
V_{\text{eff}}(\{H^{(i)}\})=m_{H_i}^2 H^{(i)\dagger}H^{(i)}
-\frac{N}{8\pi^2}\int_0^{\Lambda^2} k^2\,dk^2\:
\log\det(k^2+A),
\end{equation}
with

\begin{equation}\label{A5}
A=
\begin{pmatrix}
 g^{(i)\dagger}g^{(j)} H^{(i)\dagger}H^{(j)} &
 g^{(i)\dagger}h^{(j)} \epsilon^{ab} H^{a(i)*}H^{b(j)*} \\
-h^{(i)\dagger}g^{(j)} \epsilon^{ab} H^{a(i)}H^{b(j)} &
 h^{(i)\dagger}h^{(j)} H^{(j)\dagger}H^{(i)}
\end{pmatrix}.
\end{equation}
where summation over the indices $i$ and $j$ is understood.
The effective potential is of course gauge invariant.

\section{Change of Auxiliary Field Bases}\label{app:auxfield_bases}

In this appendix we define new bases for the auxiliary fields in the neutral and charged sectors.
We choose the new bases in such a way that the Goldstone bosons become basis vectors.
The advantage is that in each sector, neutral and charged, one of the basis vectors of the two-point proper-vertex matrix is already an eigenvector with associated eigenvalue equal to zero, i.e. a pole of the propagator.

The Goldstone theorem tells us how to express the Goldstone fields as a function of the scalar fields.
They are given by the infinitesimal displacements of the vacuum under the transformation generated by the broken generators.
The charged and neutral Goldstone boson fields are given by

\begin{eqnarray}\label{}
G^{\pm}  \propto & v\:\phi^{\pm (1)}+w\: e^{i\eta}\:\phi^{\pm (2)},\label{hGb7}\\
G \:\: \propto & \mathcal{I}m(\,v\:\phi^{0(1)}+w \:e^{-i\eta}\:\phi^{0(2)}),\label{hGb8}
\end{eqnarray}
where the fields $\phi^{0(i)}$, $\phi^{\pm(i)}$ are components of the Higgs fields $H^{(i)}=(\phi^{0(i)},\phi^{-(i)})^T$.

In the four-dimensional neutral sector we define the new basis by

\begin{equation}\label{h58}
\begin{pmatrix} \varphi^1 \\ \varphi^2 \\ \varphi^3 \\ G
\end{pmatrix}
= R\;
\begin{pmatrix}
\mathcal{R}e\,\phi^{0(1)}\\
\mathcal{I}m\,\phi^{0(1)}\\
\mathcal{R}e\,\phi^{0(2)}\\
\mathcal{I}m\,\phi^{0(2)}
\end{pmatrix},
\end{equation}
where the orthogonal transformation matrix $R$ is given by

\begin{equation}\label{h59}
R=\frac{1}{\sqrt{v^2+w^2}}
\begin{pmatrix}
w & 0 & -v\,\cos\eta & -v\,\sin\eta \\
0 & w &  v\,\sin\eta & -v\,\cos\eta \\
v & 0 &  w\,\cos\eta &  w\,\sin\eta\\
0 & v & -w\,\sin\eta &  w\,\cos\eta
\end{pmatrix}.
\end{equation}
The field $G$ is the normalized Goldstone boson given in eq. (\ref{hGb8}).
In this new basis the mass term of the neutral bosonic fields is given by

\begin{equation}\label{h60}
- \sum_{i=1,2} m_{H_i}^2\; H^{(i)\dagger}H^{(i)}\supset
-\frac{1}{2}
\begin{pmatrix}   \varphi^1, \varphi^2 , \varphi^3 , G
\end{pmatrix}
\mathcal{M}
\begin{pmatrix} \varphi^1\\ \varphi^2 \\ \varphi^3 \\ G
\end{pmatrix},
\end{equation}
with

\begin{equation}\label{h61}
\mathcal{M}=\frac{2}{v^2+w^2}
\begin{pmatrix}
w^2 m_{H_1}^2 + v^2 m_{H_2}^2 &0& v w (m_{H_1}^2-m_{H_2}^2) &0\\
0& w^2 m_{H_1}^2 + v^2 m_{H_2}^2 &0& v w (m_{H_1}^2-m_{H_2}^2)\\
v w (m_{H_1}^2-m_{H_2}^2) &0& v^2 m_{H_1}^2 + w^2 m_{H_2}^2 &0\\
0& v w (m_{H_1}^2-m_{H_2}^2) &0& v^2 m_{H_1}^2 + w^2 m_{H_2}^2
\end{pmatrix}.
\end{equation}

Now we turn to the charged sector formed by two charged fields.
The new basis is defined by

\begin{equation}\label{h70}
\begin{pmatrix}
\varphi^\pm \\ G^\pm
\end{pmatrix}
=\frac{1}{\sqrt{v^2+w^2}}
\begin{pmatrix}
w\, e^{i\eta} & -v \\
v & w\, e^{-i\eta}
\end{pmatrix}
\begin{pmatrix}
\phi^{\pm(1)} \\ \phi^{\pm(2)}
\end{pmatrix} ,
\end{equation}
where the charged field $G^\pm$ is the normalized charged Goldstone boson given in eq. (\ref{hGb7}).
In this new basis the mass term for the charged scalar fields is given by

\begin{eqnarray}\label{h71}
\lefteqn{- \sum_{i=1,2} m_{H_i}^2\; H^{(i)\dagger}H^{(i)}\supset}\\
& &-\frac{1}{v^2+w^2}
\begin{pmatrix} \varphi^+ & G^+ \end{pmatrix}
\begin{pmatrix}
w^2 m_{H_1}^2+v^2 m_{H_2}^2        &  vw\, e^{i\eta} (m_{H_1}^2-m_{H_2}^2) \\
vw\, e^{-i\eta} (m_{H_1}^2-m_{H_2}^2) &  v^2 m_{H_1}^2 + w^2 m_{H_2}^2
\end{pmatrix}
\begin{pmatrix} \varphi^- \\ G^- \end{pmatrix}. \nonumber
\end{eqnarray}

We also give the Yukawa couplings, eq. (\ref{A2}), in these bases.
The neutral boson interaction terms are given in the fermion mass basis by

\begin{equation}\label{}
\begin{split}
\mathcal{L}_{\text{Yukawa-neutral}} &=
-\frac{\varphi^1}{\sqrt{v^2+w^2}}\:
    \bar{u}^{'}(K_u\,\projectorr + K_u^\dagger\,\projectorl)u^{'}
-\frac{\varphi^2}{\sqrt{v^2+w^2}}\:
    \bar{u}^{'}(i K_u\,\projectorr -i K_u^\dagger\,\projectorl)u^{'}  \\
&\quad\quad
-\frac{\varphi^3}{\sqrt{v^2+w^2}}\:
    \bar{u}^{'}\sqrt{2}M_u u^{'}
-\frac{G}{\sqrt{v^2+w^2}}\:
    \bar{u}^{'}\sqrt{2}M_u\, i\gamma_5\, u^{'}+\;\dots,
\end{split}
\end{equation}
where the dots represent analogous terms for the down sector.
For the charged sector we found

\begin{equation}\label{}
\begin{split}
\mathcal{L}_{\text{Yukawa-charged}} &=
-\frac{\varphi^+\; e^{i\eta}}{\sqrt{v^2+w^2}}\: \bar{u}^{'}
    (K_u^\dagger\, V_{CKM}\,\projectorl -V_{CKM} K_d^\dagger\,\projectorr)d^{'} \\
&\quad
-\frac{G^+}{\sqrt{v^2+w^2}}\: \bar{u}^{'}
    (\sqrt{2}M_u V_{CKM}\,\projectorl -V_{CKM}\sqrt{2} M_d\,\projectorr)d^{'}+h.c.,
\end{split}
\end{equation}
where $\projectorl$ and $\projectorr$ are the left and right projectors.
The matrices $V_{CKM}$, $M_q$, and $K_q$ are $3\times 3$ in flavor space.
$V_{CKM}=U^{u\dagger} U^d$ is the CKM matrix, the matrices $M_q$ with $q=u,d$ are the diagonal fermion mass matrices from the up and down sector, and the matrices $K_q$ are given by

\begin{equation}
K_q\equiv w\,\lambda_q^{(1)}-v\,e^{i\eta}\lambda_q^{(2)},
\end{equation}
with $q=u,d$.

\section{Two-point Functions for the Third-generation Case}\label{app:two-point_1fam}

In this appendix we calculate the masses of the neutral and charged composite Higgs for the case  of one family of quarks and $G_{tb} \neq 0$ considered in section \ref{secc:1fam}.
We do the calculation in the auxiliary field bases defined in appendix \ref{app:auxfield_bases}.

\begin{figure}[tbp]
\begin{center}
\psfig{file=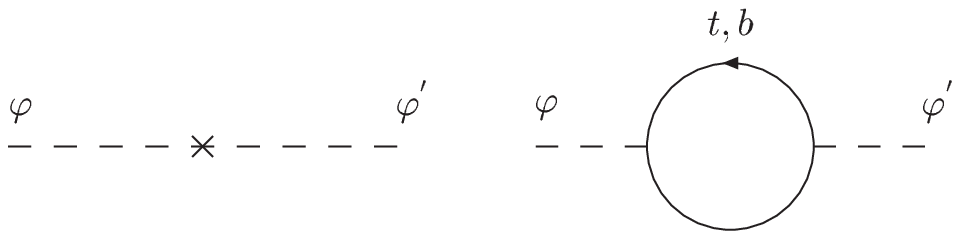,width = 8.0cm}
\end{center}
\caption{Feynman diagrams contributing to the neutral two-point proper vertices of auxiliary fields in the case of one quark family. The fields $\varphi$ and $\varphi^{'}$ stand for the four bosonic fields $\varphi^1$, $\varphi^2$, $\varphi^3$ and $G$.}
\label{fig:2hd_1fam_neutral}
\end{figure}

Let us first calculate the composite Higgs masses in the neutral sector.
As we saw in section \ref{secc:1fam} the theory is $CP$-invariant.
As a consequence the $CP$-even ($\varphi^1$ and $\varphi^3$) and
the $CP$-odd ($\varphi^2$ and $G$) fields do not mix and the $4\times 4$ two-point proper vertex matrix is a block diagonal matrix.
The Feynman diagrams we have to calculate are shown in Fig. \ref{fig:2hd_1fam_neutral}.
Using the first derivative conditions in the calculation,
the non-vanishing two-point functions are

\begin{eqnarray}
i\Gamma_{\varphi^2,\varphi^2}(p^2)&=&\frac{i N}{8\pi^2(v^2+w^2)} \Bigg\{
        \;p^2 \sum_{q=t,b} K_q^2\,I(m_q^2;p^2)  \nonumber\\
 & &    -2\Lambda^2\sum_{q=t,b}\left(1-\frac{m_q^2}{\Lambda^2}\log(\Lambda^2/m_q^2)\right)
        \left(\frac{w^3}{v}+\frac{v^3}{w} +2vw\right)
        \lambda_q^{(1)}\lambda_q^{(2)}\cos\eta  \Bigg\},    \label{h64}  \\
i\Gamma_{G,G}(p^2)          &=& \frac{i N}{8\pi^2(v^2+w^2)}
      \; p^2 \sum_{q=t,b}\: 2 \: m_q^2\: I(m_q^2;p^2),    \label{h65}     \\
i\Gamma_{\varphi^2,G}(p^2)  &=& \frac{i N}{8\pi^2(v^2+w^2)}
      \; p^2 \sum_{q=t,b} \sqrt{2}\: m_q\:
     K_q\, I(m_q^2;p^2),   \label{h66}
\end{eqnarray}
in the $CP$-odd sector, and

\begin{eqnarray}
i\Gamma_{\varphi^1,\varphi^1}(p^2)&=&\frac{i N}{8\pi^2(v^2+w^2)} \Bigg\{
    \;\sum_{q=t,b}\: (p^2-4 m_q^2)\: K_q^2\: I(m_q^2;p^2)  \nonumber \\
 & &    -2\Lambda^2\sum_{q=t,b}\left(1-\frac{m_q^2}{\Lambda^2}\log(\Lambda^2/m_q^2)\right)
        \left(\frac{w^3}{v}+\frac{v^3}{w} +2vw\right)
        \lambda_q^{(1)}\lambda_q^{(2)}\cos\eta \Bigg\},  \label{h67}   \\
i\Gamma_{\varphi^3,\varphi^3}(p^2) &=& \frac{i N}{8\pi^2(v^2+w^2)}
       \sum_{q=t,b}\: 2 \: m_q^2\:(p^2-4 m_q^2)\: I(m_q^2;p^2),    \label{h68}     \\
i\Gamma_{\varphi^1,\varphi^3}(p^2) &=& \frac{i N}{8\pi^2(v^2+w^2)}
    \:\sum_{q=t,b} \sqrt{2}\: m_q\:(p^2-4 m_q^2)
     K_q\, I(m_q^2;p^2),  \label{h69}
\end{eqnarray}
in the $CP$-even sector.
The integral $I(m_q^2;p^2)$ is given by

\begin{equation}\label{h69b}
I(m^2;p^2)=
\frac{16 \pi^2}{i} \int \frac{d^4l}{(2\pi)^4}\:
\frac{1}{(l^2-m^2)[(l+p)^2-m^2]}.
\end{equation}
The masses of the bound states are given by the values of $p^2$ at which the proper vertex matrix has vanishing eigenvalues.
From eqs. (\ref{h65}) and (\ref{h66}) we see that at $p^2=0$
a zero eigenvalue  with associated eigenvector $G$, the neutral Goldstone boson, appears.
In the $CP$-even sector the $(\varphi^1,\varphi^1)$ entry of the $2\times 2$ matrix is of order\footnote{If the factor of $\Lambda^2$ in this expression were zero, we had $G_{tb}=0$.}
$\Lambda^2$ and therefore, for $p^2\ll\Lambda^2$,
much bigger than the other matrix elements.
In first approximation the smaller eigenvalue of this matrix is given by
$i\Gamma_{\varphi^3,\varphi^3}(p^2)$.
From eq. (\ref{h68}) we see that the propagator has a pole at $p^2\approx (2 m_t)^2$.
The other two eigenvalues,
associated with a $CP$-even and a $CP$-odd field,
are of order $\Lambda^2$. \\

\begin{figure}[tbp]
\begin{center}
\psfig{file=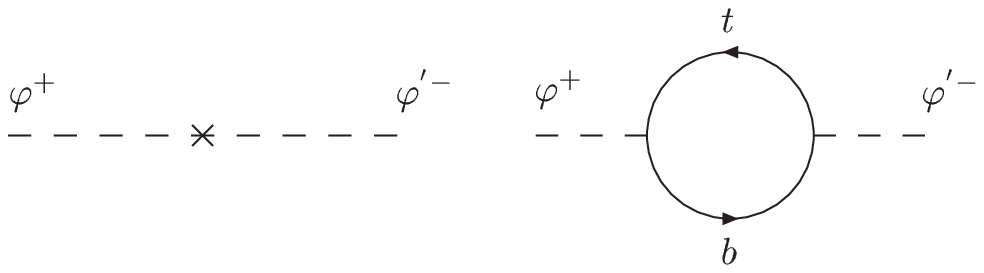,width = 8.0cm}
\end{center}
\caption{Feynman diagrams contributing to the charged two-point proper vertices of auxiliary fields in the case of one quark family. The fields $\varphi^\pm$ and $\varphi^{'\pm}$ denote the charged fields $\varphi^\pm$ and $G^\pm$.}
\label{fig:2hd_1fam_charged}
\end{figure}

Now we treat the charged sector in an analogous way.
For the two-point proper vertices,
represented by the diagrams of Fig. \ref{fig:2hd_1fam_charged},
we obtain

\begin{equation}\label{h74}
\begin{split}
i\Gamma_{\varphi^+,\varphi^-}(p^2)&=
-\frac{iN\Lambda^2}{8\pi^2 (v^2+w^2)}
\sum_{q=t,b}\left(1-\frac{m_q^2}{\Lambda^2}\log(\Lambda^2/m_q^2)\right)
\left(\frac{w^3}{v}+\frac{v^3}{w} +2vw\right)
\lambda_q^{(1)}\lambda_q^{(2)}\cos\eta            \\
& \quad+\frac{iN}{16\pi^2 (v^2+w^2)}  \Big\{
    [(p^2-m_t^2 -m_b^2)(K_t^2+K_b^2)+ 4 m_t m_b K_t K_b]\, I(m_t^2,m_b^2;p^2)      \\
& \qquad\qquad\qquad\qquad\quad
   +(K_t^2-K_b^2)
   \left(m_t^2 \log(\Lambda^2/m_t^2)-m_b^2 \log(\Lambda^2/m_b^2)\right) \Big\},
\end{split}
\end{equation}

\begin{equation}\label{h75}
\begin{split}
i\Gamma_{\varphi^+,G^-}(p^2)&=
\frac{\sqrt{2}\,iN}{16\pi^2 (v^2+w^2)}\Big\{
    [p^2(m_t K_t+m_b K_b)-(m_t K_t-m_b K_b) (m_t^2-m_b^2)]\, I(m_t^2,m_b^2;p^2)\\
&\qquad\qquad\qquad\qquad\qquad  +(m_t K_t-m_b K_b)
    \left(m_t^2 \log(\Lambda^2/m_t^2)-m_b^2 \log(\Lambda^2/m_b^2)\right) \Big\}
\end{split}
\end{equation}

\begin{equation}\label{h76}
\begin{split}
i\Gamma_{G^+,G^-}(p^2)&=
\frac{iN}{8\pi^2 (v^2+w^2)}  \Big\{
    [p^2(m_t^2+m_b^2)-(m_t^2-m_b^2)^2]\, I(m_t^2,m_b^2;p^2)      \\
& \qquad\qquad\qquad\qquad +(m_t^2-m_b^2)
   \left(m_t^2 \log(\Lambda^2/m_t^2)-m_b^2 \log(\Lambda^2/m_b^2)\right) \Big\},
\end{split}
\end{equation}
with

\begin{equation}\label{h77}
I(m_t^2,m_b^2;p^2)=
\frac{16 \pi^2}{i} \int \frac{d^4l}{(2\pi)^4}\:
\frac{1}{(l^2-m_t^2)[(l+p)^2-m_b^2]}.
\end{equation}
We have here a similar situation as in the neutral $CP$-odd sector.
The $(\varphi^+,\varphi^-)$ element is of order $\Lambda^2$ and
the other elements of the matrix vanish at $p^2=0$.
Therefore the two poles are located at $p^2=0$ and at $p^2=\mathcal{O}(\Lambda^2)$.

\section{Definitions used in section \ref{secc:3fam}}\label{app:some_definitions}

\begin{equation}\label{}
\begin{split}
\Sigma^{(0)}&=
\begin{pmatrix}
\lambda_u^{(1)\dagger}\lambda_u^{(2)} & 0 \\
0 & \lambda_d^{(2)}\lambda_d^{(1)\dagger}
\end{pmatrix},\\
\Sigma^{(i)}&=
\begin{pmatrix}
\lambda_u^{(i)\dagger}\lambda_u^{(i)} & 0 \\
0 & \lambda_d^{(i)}\lambda_d^{(i)\dagger}
\end{pmatrix}\:,
\;\qquad\text{for }i=1,2.
\end{split}
\end{equation}

\begin{equation}\label{}
\begin{split}
a&=\text{tr}\big[
              \:\mathcal{R}e(W^u M_u U^{u\dagger})
              \:\mathcal{I}m(U^u M_u W^{u\dagger}) +
              \:\mathcal{I}m(W^d M_d U^{d\dagger})
              \:\mathcal{R}e(U^d M_d W^{d\dagger})\: \big]\\
&\qquad\qquad -\sum_{i=u,c,t}\frac{m_i^2}{\Lambda^2}\log(\Lambda^2/m_i^2)\:
                 \mathcal{R}e\Big(\:
  W^{u\dagger}\:\mathcal{R}e(W^u M_u U^{u\dagger})
              \:\mathcal{I}m(U^u M_u W^{u\dagger})\:W^u\Big)_{ii}     \\
&\qquad\qquad -\sum_{i=d,s,b}\frac{m_i^2}{\Lambda^2}\log(\Lambda^2/m_i^2)\:
                 \mathcal{R}e\Big(\:
  W^{d\dagger}\:\mathcal{I}m(W^d M_d U^{d\dagger})
              \:\mathcal{R}e(U^d M_d W^{d\dagger})\:W^d\Big)_{ii},
\end{split}
\end{equation}

\begin{equation}\label{}
\begin{split}
b&=\text{tr}\big[
              \:\mathcal{I}m(W^u M_u U^{u\dagger})
              \:\mathcal{I}m(U^u M_u W^{u\dagger})\: +
              \:\mathcal{I}m(W^d M_d U^{d\dagger})
              \:\mathcal{I}m(U^d M_d W^{d\dagger})\:\big]\\
&\qquad\qquad -\sum_{i=u,c,t}\frac{m_i^2}{\Lambda^2}\log(\Lambda^2/m_i^2)\:
                 \mathcal{R}e\Big(\:
  W^{u\dagger}\:\mathcal{I}m(W^u M_u U^{u\dagger})
              \:\mathcal{I}m(U^u M_u W^{u\dagger})\:W^u\Big)_{ii}     \\
&\qquad\qquad -\sum_{i=d,s,b}\frac{m_i^2}{\Lambda^2}\log(\Lambda^2/m_i^2)\:
                 \mathcal{R}e\Big(\:
  W^{d\dagger}\:\mathcal{I}m(W^d M_d U^{d\dagger})
              \:\mathcal{I}m(U^d M_d W^{d\dagger})\:W^d\Big)_{ii},
\end{split}
\end{equation}

\begin{equation}\label{}
\begin{split}
c&=-\sum_{i=u,c,t}\frac{m_i^2}{\Lambda^2}\log(\Lambda^2/m_i^2)\:
                 \mathcal{I}m\Big(\:
  W^{u\dagger}\:\mathcal{R}e(W^u M_u U^{u\dagger})
              \:\mathcal{I}m(U^u M_u W^{u\dagger})\:W^u\Big)_{ii}     \\
&\qquad  -\sum_{i=d,s,b}\frac{m_i^2}{\Lambda^2}\log(\Lambda^2/m_i^2)\:
                 \mathcal{I}m\Big(\:
  W^{d\dagger}\:\mathcal{I}m(W^d M_d U^{d\dagger})
              \:\mathcal{R}e(U^d M_d W^{d\dagger})\:W^d\Big)_{ii},
\end{split}
\end{equation}

\begin{equation}\label{}
\begin{split}
T&=\frac{v^{'}}{\sqrt{2}}\;
 \Big(\lambda_u^{(2)\dagger}\,V_{CKM}\,\lambda_d^{(1)\dagger}-
      \lambda_u^{(1)\dagger}\,V_{CKM}\,\lambda_d^{(2)\dagger}\Big),\\
I(m_i^2,m_j^2)&=
\int_0^{\Lambda^2}\frac{k^2dk^2}{(k^2+m_i^2)(k^2+m_j^2)},
\end{split}
\end{equation}

\begin{equation}\label{3f23}
\begin{split}
\frac{\partial A}{\partial v^{'}}\Big|_{(v,w,\eta,0)} &=
v\: \Sigma^{(1)}
+\frac{w}{2}\left(
\Sigma^{(0)} e^{i\eta}+\Sigma^{(0)\dagger} e^{-i\eta}\right) ,\\
\frac{\partial A}{\partial w^{'}}\Big|_{(v,w,\eta,0)} &=
w\: \Sigma^{(2)}
+\frac{v}{2}\left(
\Sigma^{(0)} e^{i\eta}+\Sigma^{(0)\dagger} e^{-i\eta}\right) , \\
\frac{\partial A}{\partial \eta^{'}}\Big|_{(v,w,\eta,0)} &=
\frac{ivw}{2}\left(
\Sigma^{(0)} e^{i\eta}-\Sigma^{(0)\dagger} e^{-i\eta}\right).
\end{split}
\end{equation}

\section{Two-point Functions for the Three Family Case}\label{app:two-point_3fam}

In this appendix we calculate the composite-field two-point functions of the three family case considered in section \ref{secc:3fam}.
The situation is analogous to the one of appendix \ref{app:two-point_1fam} where only one family of quarks was considered.
The values of the flowing momentum for which an eigenvalue of the proper vertex matrix becomes zero, correspond to the composite Higgs masses we are looking for.
We use the neutral and charged auxiliary field bases defined in appendix \ref{app:auxfield_bases}.
In the calculation we also apply the first derivative conditions from the minimization of the effective potential.
The non-vanishing neutral proper vertices are given by

\begin{equation}\label{2pf_neutral1}
\begin{split}
i\Gamma_{(\varphi^1,\varphi^1)/(\varphi^2,\varphi^2)}&(p^2)
 = \frac{i N}{8\pi^2(v^2+w^2)} \Bigg\{
    \; \frac{4\, c\, \Lambda^2}{\sin^2\eta}
        \left(\frac{w^2}{v^2}+\frac{v^2}{w^2} +2\right)   \\
& +\sum_{i=1}^6\: m_i^2\log\Lambda^2/m_i^2
    \big[ w^2(\Sigma_{ii}^{(1)}-\tilde{\Sigma}_{ii}^{(1)})
          +v^2(\Sigma_{ii}^{(2)}-\tilde{\Sigma}_{ii}^{(2)})     \\
&  \qquad\qquad\qquad\qquad\qquad\qquad
 -2vw \,\mathcal{R}e((\Sigma_{ii}^{(0)}-\tilde{\Sigma}_{ii}^{(0)})e^{i\eta})
      \big]                                                    \\
&  + \sum_{r,s=1}^3 \sum_{q=u,d}\: I(m_r^2,m_s^2;p^2)
 \Big[(p^2-m_r^2-m_s^2) (K_q)_{rs}(K_q^\dagger)_{sr}         \\
&  \qquad\qquad\qquad\qquad
   \mp m_r m_s ((K_q)_{rs}(K_q)_{sr}+(K_q^\dagger)_{rs}(K_q^\dagger)_{sr})
 \Big] \Bigg\},
\end{split}
\end{equation}
where the minus sign in the last term corresponds to $i\Gamma_{\varphi^1,\varphi^1}$ and the plus sign to $i\Gamma_{\varphi^2,\varphi^2}$.
The matrices $\tilde{\Sigma}^{(i)}$ are defined by

\begin{equation}\label{}
\begin{split}
\tilde{\Sigma}^{(0)}&=
\begin{pmatrix}
\lambda_u^{(2)}\lambda_u^{(1)\dagger} & 0 \\
0 & \lambda_d^{(1)\dagger}\lambda_d^{(2)}
\end{pmatrix},\\
\tilde{\Sigma}^{(i)}&=
\begin{pmatrix}
\lambda_u^{(i)}\lambda_u^{(i)\dagger} & 0 \\
0 & \lambda_d^{(i)\dagger}\lambda_d^{(i)}
\end{pmatrix}\:,
\;\qquad\text{for }i=1,2.
\end{split}
\end{equation}
Further neutral proper vertices are given by

\begin{figure}[tbp]
\begin{center}
\psfig{file=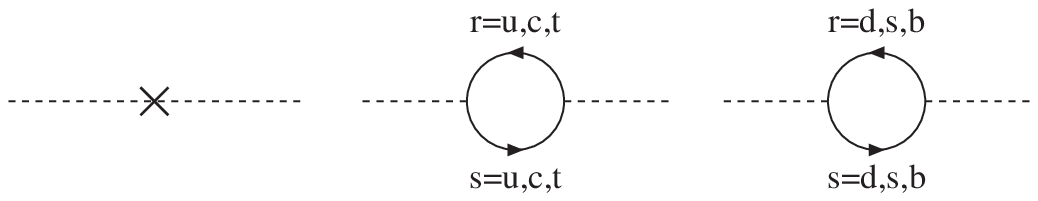,width = 9.9cm}
\end{center}
\caption{Diagrammatic representation of the contributions to the neutral two-point proper vertices.}
\label{fig:3fam_masses_neutral}
\end{figure}

\begin{eqnarray}
i\Gamma_{\varphi^1,\varphi^2}(p^2)&=&
 \frac{2i N}{8\pi^2(v^2+w^2)} \sum_{r,s=1}^3 \sum_{q=u,d}\:
 m_r m_s \, \mathcal{I}m[(K_q)_{rs}(K_q)_{sr}]\, I(m_r^2,m_s^2;p^2),
                                                    \label{2pf_neutral2}    \\
i\Gamma_{\varphi^3,\varphi^3}(p^2)&=&
 \frac{2i N}{8\pi^2(v^2+w^2)} \sum_{i=1}^6 \:
 (p^2-4m_i^2)m_i^2\, I(m_i^2,m_i^2;p^2)  ,       \label{2pf_neutral3}          \\
i\Gamma_{G,G}(p^2)&=&
 \frac{2i N}{8\pi^2(v^2+w^2)}\,p^2\, \sum_{i=1}^6 \:
 m_i^2\, I(m_i^2,m_i^2;p^2)    ,                   \label{2pf_neutral4}      \\
i\Gamma_{\varphi^1,\varphi^3}(p^2)&=&
 \frac{i N}{8\pi^2(v^2+w^2)} \sum_{i=1}^6 \:
 (p^2-4m_i^2)\,\sqrt{2}\,m_i\,\mathcal{R}e[(K_{u/d})_{ii}]\, I(m_i^2,m_i^2;p^2) ,
                                                   \label{2pf_neutral5}  \\
i\Gamma_{\varphi^2,G}(p^2)&=&
 \frac{i N}{8\pi^2(v^2+w^2)}\,p^2\, \sum_{i=1}^6 \:
 \sqrt{2}\,m_i\,\mathcal{R}e[(K_{u/d})_{ii}]\, I(m_i^2,m_i^2;p^2)  ,
                                                      \label{2pf_neutral6}   \\
i\Gamma_{\varphi^1,G}(p^2)&=&
 \frac{i N}{8\pi^2(v^2+w^2)}\,p^2\, \sum_{i=1}^6 \:
 \sqrt{2}\,m_i\,\mathcal{I}m[(K_{u/d})_{ii}]\, I(m_i^2,m_i^2;p^2),
                                                      \label{2pf_neutral7}   \\
i\Gamma_{\varphi^2,\varphi^3}(p^2)&=&
 \frac{i N}{8\pi^2(v^2+w^2)} \sum_{i=1}^6 \:
 (4m_i^2-p^2)\,\sqrt{2}\,m_i\,\mathcal{I}m[(K_{u/d})_{ii}]\, I(m_i^2,m_i^2;p^2) ,   \label{2pf_neutral8}
\end{eqnarray}
where in Eqs. (\ref{2pf_neutral5})-(\ref{2pf_neutral8}) $K_{u/d}$ stands for $K_u$ for $i=1,2,3$ and for $K_d$ for $i=4,5,6$.
The relevant diagrams of the neutral sector are shown in Fig. \ref{fig:3fam_masses_neutral}.

\begin{figure}[tbp]
\begin{center}
\psfig{file=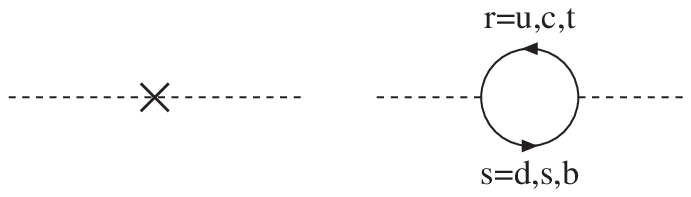,width = 6.6cm}
\end{center}
\caption{Diagrammatic representation of the contributions to the charged two-point proper vertices.}
\label{fig:3fam_masses_charged}
\end{figure}

In the charged sector the two-point proper vertices are given by

\begin{equation}\label{2pf_charged1}
\begin{split}
i\Gamma_{\varphi^+,\varphi^-}(p^2)
&= \,\frac{2i N}{8\pi^2(v^2+w^2)} \frac{c\,\Lambda^2}{\sin^2\eta}
      \left(\frac{w^2}{v^2}+\frac{v^2}{w^2} +2\right)             \\
&  -\frac{i N}{16\pi^2(v^2+w^2)}
 \sum_{\begin{matrix}\scriptstyle r=u,c,t \\
                     \scriptstyle s=d,s,b   \end{matrix}} \, \Big\{
 (K_u^\dagger V_{CKM})_{rs} (V_{CKM}^\dagger K_u)_{sr}
        \,\big[ J(m_s^2,m_r^2;p^2) + 2 m_s^2         \big]     \\
&\qquad\qquad\qquad\qquad\qquad
+(V_{CKM} K_d^\dagger)_{rs} (K_d V_{CKM}^\dagger)_{sr}
        \,\big[ J(m_r^2,m_s^2;p^2) + 2 m_r^2         \big]     \\
&\qquad-\big[ (K_u^\dagger V_{CKM})_{rs} (K_d V_{CKM}^\dagger)_{sr}
        +(V_{CKM} K_d^\dagger)_{rs} (V_{CKM}^\dagger K_u)_{sr} \big]
  2 m_r m_s\, I(m_r^2,m_s^2;p^2)   \Big\},
\end{split}
\end{equation}

\begin{equation}\label{2pf_charged2}
i\Gamma_{G^+,G^-}(p^2)
= \frac{-i N}{8\pi^2(v^2+w^2)}
 \sum_{\begin{matrix}\scriptstyle r=u,c,t \\
                     \scriptstyle s=d,s,b   \end{matrix}}
 \;(V_{CKM})_{rs} (V_{CKM}^\dagger)_{sr} \,  \Big\{
      m_r^2 \, J(m_s^2,m_r^2;p^2) +m_s^2\, J(m_r^2,m_s^2;p^2)        \Big\},
\end{equation}

\begin{equation}\label{2pf_charged3}
\begin{split}
i\Gamma_{\varphi^+,G^-}(p^2)
= \frac{-\sqrt{2}\, i N e^{i\eta}}{16\pi^2(v^2+w^2)}
 \sum_{\begin{matrix}\scriptstyle r=u,c,t \\
                     \scriptstyle s=d,s,b   \end{matrix}}
  \;(V_{CKM})_{rs} \, \Big\{&m_r (V_{CKM}^\dagger K_u)_{sr}\, J(m_s^2,m_r^2;p^2)  \\
                           &+m_s (K_d V_{CKM}^\dagger)_{sr}\, J(m_r^2,m_s^2;p^2)
                      \Big\},
\end{split}
\end{equation}
where $J(m_r^2,m_s^2;p^2)$ defined by
\begin{equation}\label{2pf_charged4}
J(m_r^2,m_s^2;p^2)=m_r^2\log\Lambda^2/m_r^2 - m_s^2\log\Lambda^2/m_s^2
                 -(p^2+m_r^2-m_s^2)\, I(m_r^2,m_s^2;p^2),
\end{equation}
vanishes at $p^2=0$.
The relevant diagrams of the charged sector are shown in Fig. \ref{fig:3fam_masses_charged}.

In all the two-point proper vertices the quadratic divergences cancel.
As expected, the neutral and charged Goldstone bosons correspond to eigenvectors of the proper vertex matrix at $p^2=0$ with vanishing eigenvalues.
Besides these three Goldstone bosons there are three neutral and one charged Higgs particles.
The masses of these particles are roughly $\sim 2m_q$,
with $m_q$ a general quark mass.
We find that in this approximation one neutral Higgs mass is $\sim 2m_t$ and the rest of the Higgs masses are much smaller.

\end{appendix}

\end{document}